\documentclass[twocolumn,english,reprint]{revtex4}
\usepackage[T1]{fontenc}
\usepackage[latin9]{inputenc}
\usepackage{color}
\usepackage{textcomp}
\usepackage{amsmath}
\usepackage{graphicx}

\makeatletter

\newcommand{\lyxmathsym}[1]{\ifmmode\begingroup\def\b@ld{bold}
  \text{\ifx\math@version\b@ld\bfseries\fi#1}\endgroup\else#1\fi}

\@ifundefined{textcolor}{}
{%
 \definecolor{BLACK}{gray}{0}
 \definecolor{WHITE}{gray}{1}
 \definecolor{RED}{rgb}{1,0,0}
 \definecolor{GREEN}{rgb}{0,1,0}
 \definecolor{BLUE}{rgb}{0,0,1}
 \definecolor{CYAN}{cmyk}{1,0,0,0}
 \definecolor{MAGENTA}{cmyk}{0,1,0,0}
 \definecolor{YELLOW}{cmyk}{0,0,1,0}
 }

\@ifundefined{definecolor}
 {\usepackage{color}}{}

\usepackage{babel}

\usepackage{babel}

\usepackage{babel}

\usepackage{babel}

\usepackage{babel}

\usepackage{babel}

\makeatother

\usepackage{babel}
\begin{document}

\title{\textcolor{black}{Effective 90-degree magnetization  rotation in}
Co$_{2}$FeAl thin film/Piezoelectric system probed by microstripline
ferromagnetic resonance}

\author{M. Gueye $^{1}$}

\author{F. Zighem$^{1}$}

\email{zighem@univ-paris13.fr}

\selectlanguage{english}%

\author{M. Belmeguenai$^{1}$}

\author{M. Gabor$^{2}$}

\author{C. Tiusan$^{2}$}

\author{D. Faurie$^{1}$}

\affiliation{$^{1}$LSPM-CNRS, Université Paris XIII, Sorbonne Paris Cité, 93430
Villetaneuse, France}

\affiliation{$^{2}$Center for Superconductivity, Spintronics and Surface Science,
Technical University of Cluj-Napoca, Str. Memorandumului No. 28 RO-400114,
Cluj-Napoca, Romania}

\date{July 7$^{th}$ 2015}
\begin{abstract}
Microstripline ferromagnetic resonance technique has been used to
study the indirect magnetoelectric coupling occurring in an artificial
magnetoelectric heterostructure consisting of a magnetostrictive thin
film cemented onto a piezoelectric actuator. Two different modes (sweep-field
and sweep-frequency modes) of this technique have been employed to
quantitatively probe the indirect magnetoelectric coupling and to
observe a voltage induced magnetization rotation (of 90 degree). This
latter has been validated by the experimental frequency variation
of the uniform mode and by the amplitude of the sweep-frequency spectra.
\end{abstract}


\maketitle
In recent decades, the development and the sophistication of growth
and characterization on the atomic and nanometric scale, has led to
the elaboration of smart materials which reveal captivating phenomena.
In the midst of these materials, multiferroics magnetoelectric materials,
currently at the cutting edge of spintronics, have attracted the attention
of many groups \cite{Spaldin_Science_2005,Cheong_Nature_Materials_2007,Ramesh_Nature_Materials_2007,Bibes_Nature_Materials_2008,Eerenstein_Nature_2006}.
A flurry of research activities has been launched in order to explore
the physics behind these materials. From the applications standpoint,
an alternative approach to overcome the scientific impediments of
weak magnetoelectric coupling in single-phase multiferroics are two-phase
systems of ferromagnetic and ferroelectric constituents \cite{Srinivasan_JAP_Review_2008}.
One of the major advantages of ferromagnetic/ferroelectric extrinsic
multiferroics as reported in the literature is the flexibility in
the choice of the materials \cite{Srinivasan_JAP_Review_2008}. One
of the main idea behind the fabrication of artificial magnetoelectric
multiferroics is the electric or voltage control of the magnetization
using electric fields for low power and ultra fast next generation
electronics \cite{Nan_Advanced Materials _2011,Sun_Spin_2012}. The
straightforward heterostructure allowing such a control appears to
be the one presenting a piezoelectric/magnetostrictive interface wherein
the interaction vector is the voltage induced in-plane strains\cite{Srinivasan_JAP_Review_2008,Gueye_APL_Ni_2014,Brandlmaier_New_Journal_Physics_2009,Zighem_JAP_2013,Scott_PRB_2010,Thiele_2007,Brandlmaier_PRB_2008}. 

\begin{figure}
\includegraphics[bb=40bp 155bp 670bp 595bp,clip,width=8cm]{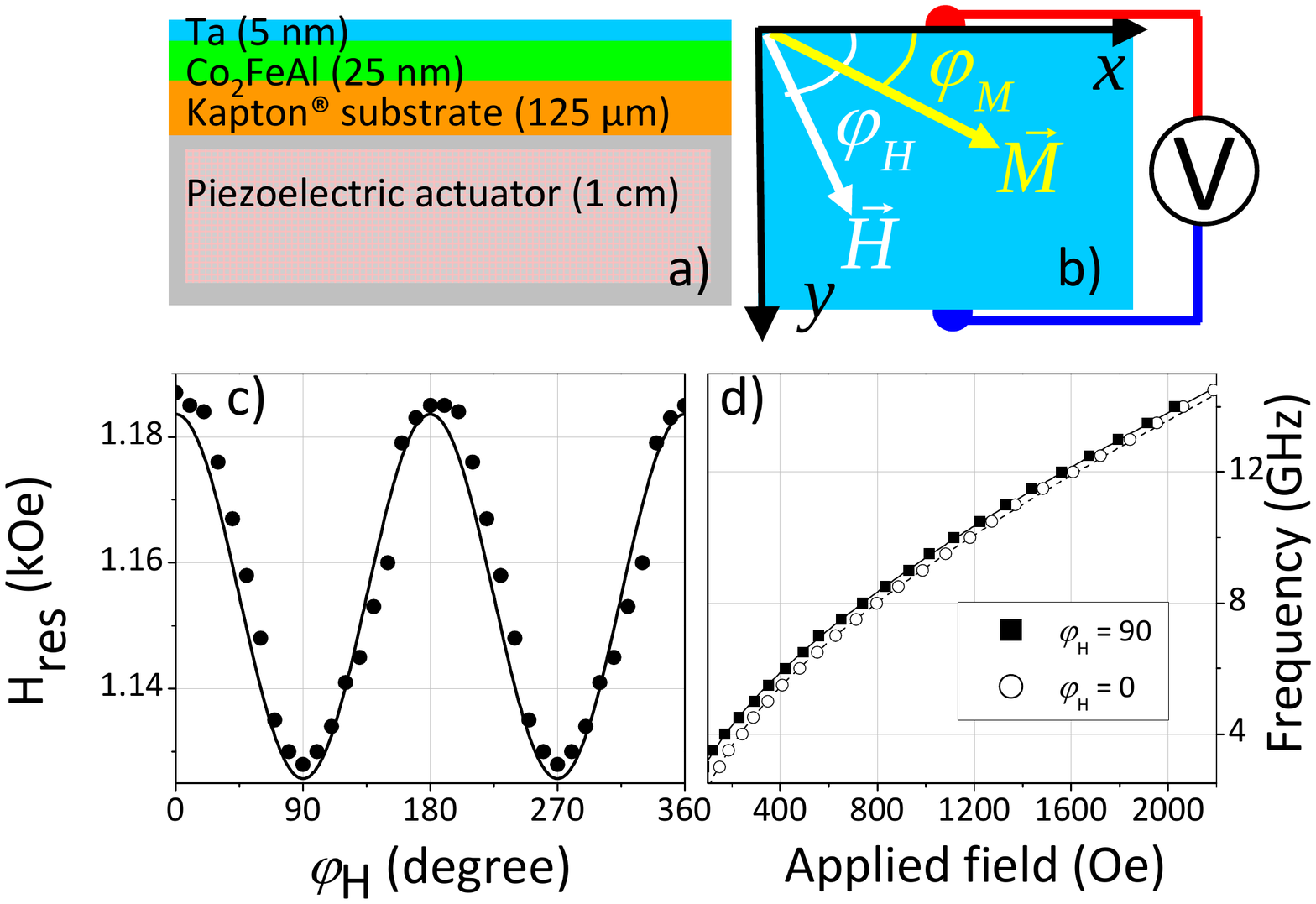}

\caption{a) Depiction of the cross section view of the investigated system
illustrating the CFA/Kapton\textregistered{} system deposited onto
a piezoelectric actuator. b) Top view sketch of the system showing
the different angles used in the text. c) Angular dependence of the
resonance field measured at 10 GHz. d) Experimental and calculated
frequency dependence of the uniform mode measured along the easy and
hard axis ($\varphi_{H}=90$ and $\varphi_{H}=0$).}

\label{Fig_Sketch_Actuator_Zero_Voltage_FMR} 
\end{figure}

The present study concerns the experimental observation of the voltage
induced magnetization rotation in an artificial magnetoelectric heterostructure
by probing the magnetic resonance of the uniform precession mode.
A method based only on microstripline ferromagnetic resonance (MS-FMR)
experiments is presented \cite{Zighem_JAP_2014}. The two different
modes of this technique are presented and used to perform a quantitative
characterization of the effective magnetoelectric coupling, as well
as of the voltage induced magnetization-rotation. Figures \ref{Fig_Sketch_Actuator_Zero_Voltage_FMR}-a)
and -b) show the artificial heterostructure composed of a 25 nm Co$_{2}$FeAl
(CFA) thin film grown onto a 125 \textmu{}m thick polyimide (Kapton\textregistered{})
flexible substrate and then cemented on a piezoelectric actuator.
The CFA film was deposited onto the Kapton\textregistered{} substrate
by rf sputtering. The deposition residual pressure was of around $4\times10^{-9}$
mbar, while the working Ar pressure was of $1.3\times10{}^{-3}$ mbar.
A 5 nm thick Ta cap layer was deposited on the top of the CFA film
in order to protect it from oxidation. CFA was chosen because of its
non-negligible magnetostriction coefficient at saturation even in
a polycrystalline film with no preferred orientation ($\lambda\approx14\times10^{-6}$)\cite{Gueye_APL_CFA_2014,Sun_CFA_JAP_2015}
which will produce strong indirect magnetoelectric field and also
because it is a serious candidate for spintronic applications \cite{Belmeguenai_PRB_2013,Wang_APL_CFA_2009,Gabor_Spin_2014}.
\textcolor{black}{Moreover, we have shown that Kapton substrate does
not affect the whole system magnetoelectric behavior because of its
high compliance \cite{Gueye_APL_Ni_2014,Zighem_JAP_2013}.}

MS-FMR experiments can be carried out in two different modes: i) by
fixing the microwave driving frequency $(f_{0})$ and sweeping the
applied magnetic field $(H)$ \cite{Zighem_JAP_2013,Zighem_JAP_2014,Belmeguenai_PRB_2013}
or ii) by fixing the applied magnetic field $(H_{0})$ and sweeping
the microw\textcolor{black}{ave frequency $(f$) \cite{Zighem_JPCM_2008}.
A resonance field ($H_{res}$ at fixed $f_{0}$) and a frequency resonance
($f_{res}$ at fixed $H_{0}$) are extracted from these two modes
each having their own advantage. Indeed, the sweep-field mode is more
sensitive than the sweep-frequency one (both modes are of course less
sensitive than standard FMR technique where resonant microwave cavities
are used). However, contrary to the sweep-field mode, the sweep-frequency
mode presents the advantage of performing spectra without disturbing
the magnetization distribution of the studied systems \cite{Labrune_PRB_2004},
provided that the rf field $\vec{h}_{rf}$ amplitude is weak enough
}(which is the case here). These two modes have been used in the present
study to characterize the indirect magnetoelectric field and to probe
the voltage induced  magnetization rotation.

\begin{figure}
\includegraphics[bb=40bp 170bp 510bp 595bp,clip,width=8cm]{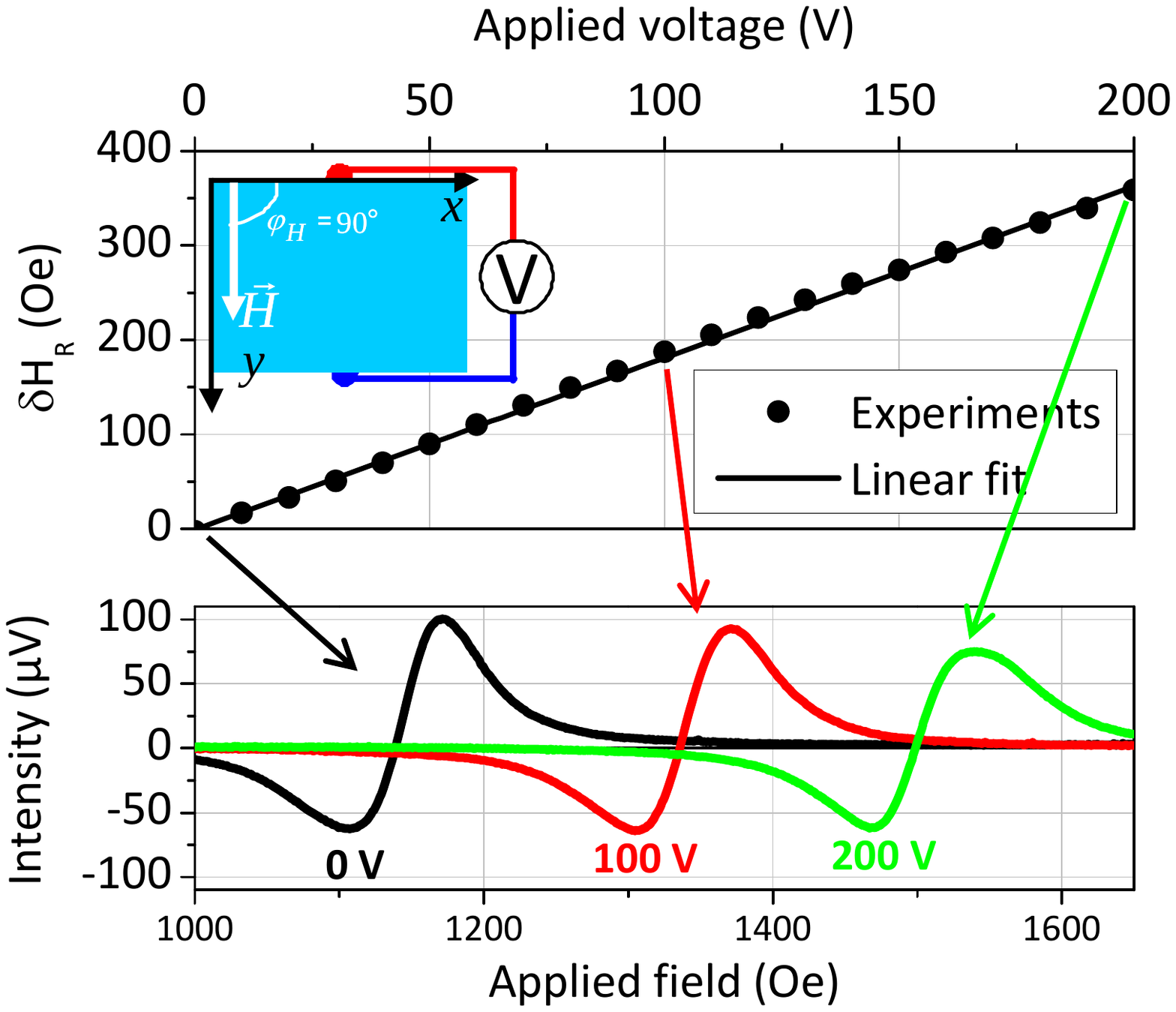}

\caption{Up-graph: Resonance field shift ($\delta H_{res}=H_{res}(V)-H_{res}(0)$)
variation as function of the applied voltage. Circles refer to experimental
data while solid line represents to the best fit. Down-graph: typical
MS-FMR spectra recorded at different applied voltage (0 V, 100 V and
200 V).}

\label{Fig_dHres} 
\end{figure}

A beforehand characterization of the magnetic properties of the heterostructure
has been performed in the benchmark state (i. e. absence of applied
voltage) in order to measure the effecti\textcolor{black}{ve magnetization
($M_{eff}$) and the gyromagnetic factor ($\gamma$) as well as the
unavoidable uniaxial anisotropy ($H_{u}$) due to in-plane non-equibiaxial
residual stresses. This initial anisotropy is attributed to imperfect
flatness of the Kapton substrate during deposition and/or cementation
on the actuator \cite{Gueye_APL_Ni_2014}. The quantitative evaluation
of} this initial uniaxial anisotropy is performed through the angular
dependence of the uniform mode resonance field (see Figure \ref{Fig_Sketch_Actuator_Zero_Voltage_FMR}-c))
measured at $f_{0}=10$ GHz \cite{Zighem_JAP_2014}. Furthermore,
one can note that this anisotropy is aligned along the $y$-axis since
the resonance field minimum reaches a minimum at $\varphi_{H}=90\text{\textdegree}$.
In addition, the frequency dependencies of the uniform mode, measured
along the easy ($\varphi_{H}=90\lyxmathsym{\textdegree}$) and hard
($\varphi_{H}=0\lyxmathsym{\textdegree}$) axes are shown in Figure
\ref{Fig_Sketch_Actuator_Zero_Voltage_FMR}-d). The full lines in
Figure \ref{Fig_Sketch_Actuator_Zero_Voltage_FMR}-c) and -d) are
the best fits to the experimental data and allowed the determination
of the following parameters: $M_{eff}=720$ emu.cm$^{-3}$, $\gamma=1.835\times10^{7}$
Hz.Oe$^{-1}$ and $H_{u}\sim30$ Oe, which are close to the values
previously obtained for similar films \cite{Gueye_APL_CFA_2014}.

\begin{figure}
\includegraphics[bb=50bp 160bp 670bp 580bp,clip,width=7.5cm]{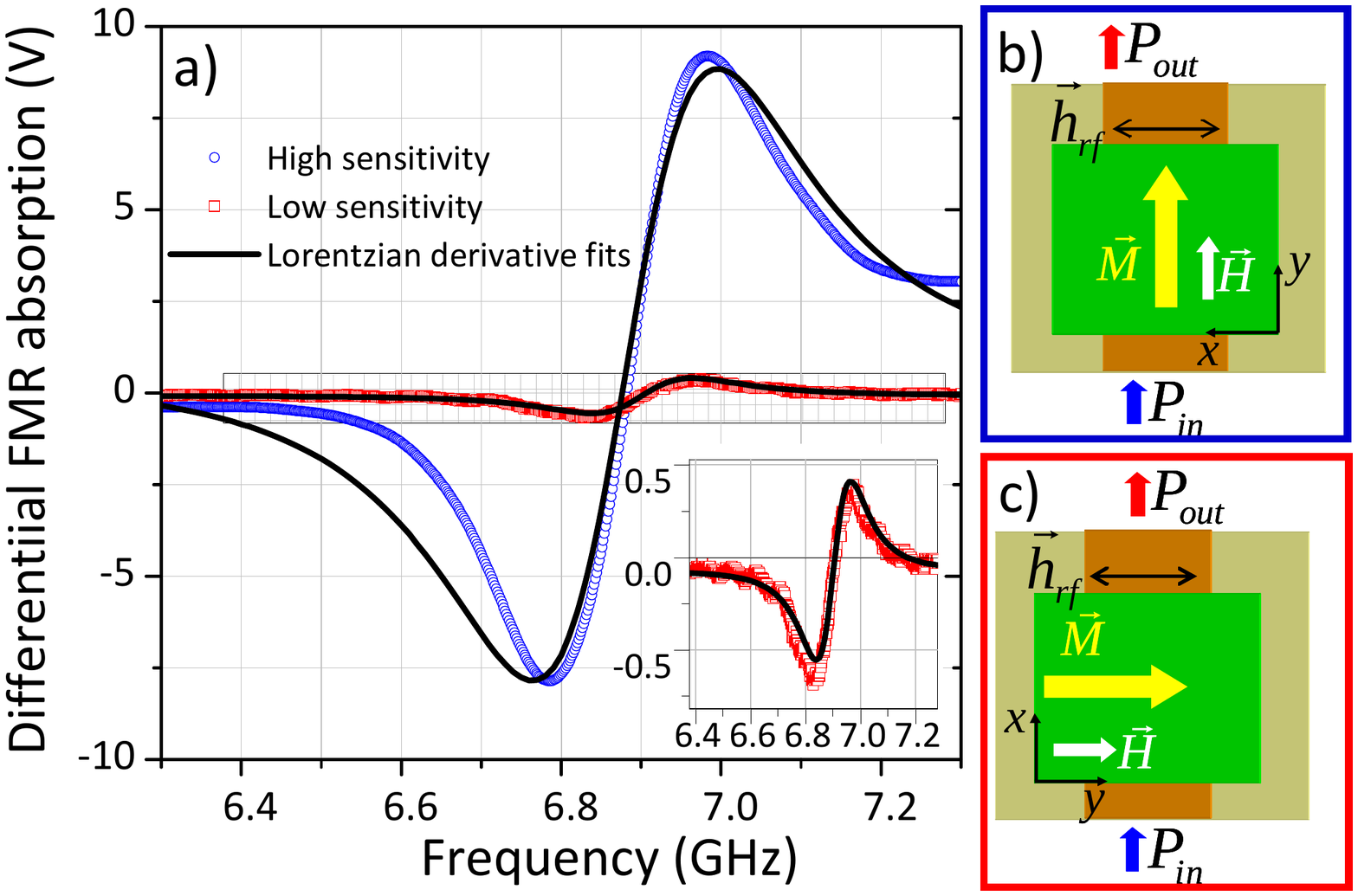}

\caption{a) Sweep-frequency FMR spectra obtained according to the two configurations
depicted in b) and c) where $P_{in}$ and $P_{out}$ are the injected
and transmitted radio frequency current. The only difference comes
from the pumping field ($\vec{h}_{rf}$) which either parallel (red
squares) or perpendicular (blue circles) to $\vec{M}$ direction giving
rise to the high and low sensitivities geometries. The inset shows
a zoom of the marked area. Note that a saturating field of 500 Oe
was applied to ensure a uniform magnetization distribution. The solid
lines show fits to the first derivative of the standard Lorentzian
curve.}

\label{Fig_Illustrative_Spectra} 
\end{figure}

Sweep-field MS-FMR spectra have been recorded as a function of the
applied voltage (for $f_{0}=10$ GHz). This latter was varied from
0 V to 200 V by 10 V steps. Figure \ref{Fig_dHres} presents typical
spectra measured at different applied voltages (0V, 100 V and 200
V) with an applied magnetic field along $y$ axis ($\varphi_{H}=90\text{\textdegree}$,
i. e. along the initial easy axis). A shift of the resonance field,
defined as $\delta H_{res}=H_{res}(V)-H_{res}(0)$, is clearly observed.
The voltage dependence of $\delta H_{res}$ is shown in Figure \ref{Fig_dHres},
where the solid line corresponds to a linear fit showing that $\delta H_{res}$
is positive and almost proportional to $V$ . This means that the
$y$ direction gradually becomes a hard magnetization axis, or a less
easy one for the magnetization direction \cite{Zighem_JAP_2013,Zighem_JAP_2014,Sun_CFA_JAP_2015}.
This is due to the positive magnetostriction coefficient of Co$_{2}$FeAl\cite{Gueye_APL_CFA_2014,Sun_CFA_JAP_2015}
and to the nearly uniaxial tensile stress (i. e. $\sigma_{xx}(V)>0$)
along the $x$ direction applied by the piezoelectric actuator to
the film (for this range of applied voltage: 0 to 200V\cite{Zighem_RSI_2014}).
Note that a non linear and hysteretic behavior is observed if the
voltage is applied backward from 200 V to 0 V (not shown here) due
to the intrinsic properties of the piezoelectric material \cite{Gueye_APL_Ni_2014,Brandlmaier_New_Journal_Physics_2009,Zighem_JAP_2013,Zighem_JAP_2014,Zighem_RSI_2014}.
It should be noted that the $\delta H_{res}(V)$ curve presented in
Figure \ref{Fig_dHres} has been measured several times and all the
obtained curves were superimposed. Furthermore, the resonance field
shift can also be viewed as an indirect magnetoelectric field $H_{me}(V)=\delta H_{res}(V)$
aligned along $x$ direction, i. e. perpendicular to the initial easy
axis induced by the deposition conditions. The evaluation of the voltage
dependence of $H_{me}$ will serve to quantitatively study the magnetization
rotation using the sweep-frequency mode of the MS-FMR setup.

The sweep-frequency-mode has been used to probe the magnetization
direction as function of the applied voltage by measuring both the
resonance frequency and the spectra amplitudes. For this purpose,
the selective rule illustrated in Figure \ref{Fig_Illustrative_Spectra}
has been be used. In this Figure, two spectra performed with a similar
applied magnetic field (of 500 Oe) along $y$ are presented; the only
difference comes from the direction of the weak rf field $\vec{h}_{rf}$
(also called pumping field) emanating from the microstripline, which
is either parallel ($\vec{h}_{rf}\parallel\vec{M}$) or perpendicular
($\vec{h}_{rf}\perp\vec{M}$) to the static magnetization. Indeed,
the efficiency of the pumping field is maximum when $\vec{h}_{rf}\perp\vec{M}$
(Figure \ref{Fig_Illustrative_Spectra}-b) and minimum when $\vec{h}_{rf}\parallel\vec{M}$
(Figure \ref{Fig_Illustrative_Spectra}-b). In the present experimental
conditions, a factor 20 is found between these two geometries. Furthermore,
Counil \textit{et al.} \cite{Counil_JAP_2004,Counil_JAP_2005} demonstrated
both theoretically and experimentally that for a pure Lorentzian profile
of the signal, the amplitude of this signal is directly proportional
to $\sin^{2}\varphi$, where $\varphi$ is the angle between the pumping
field and the static magnetization direction. Therefore, the voltage-induced
rotation of the magnetization has been probed by measuring the evolution
of the relative amplitude of the signal as function of the applied
voltage. 

\textcolor{black}{}
\begin{figure}
\textcolor{black}{\includegraphics[bb=30bp 40bp 660bp 595bp,clip,width=7.5cm]{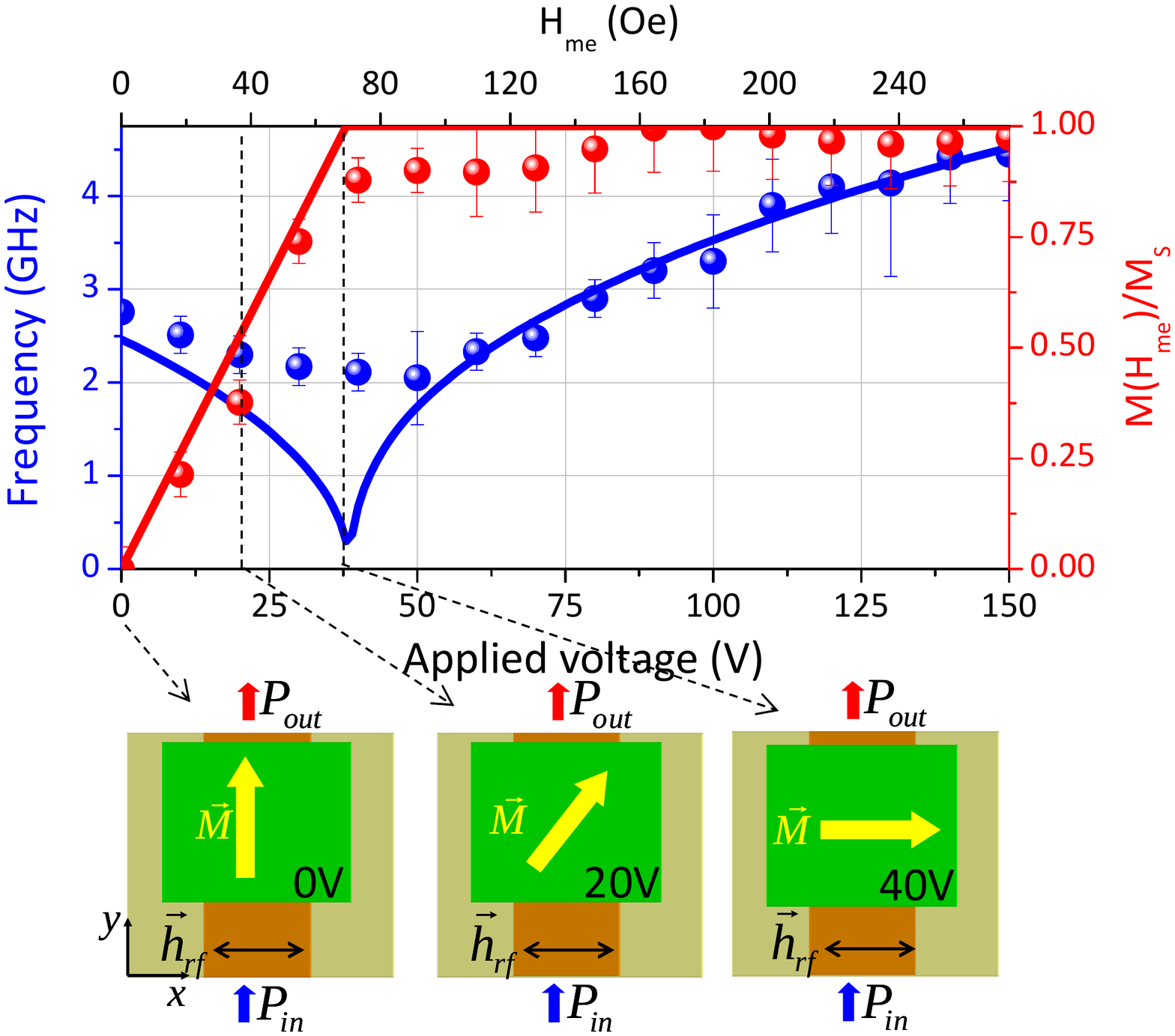}}

\textcolor{black}{\caption{Voltage dependencies of the uniform mode frequency (blue symbols)
and of the normalized $x$-component of the magnetization $m_{x}(H_{me})$
with a pumping field applied along $x$. The $m_{x}(H_{me})$ values
have been deduced from equation \ref{eq:amplitude}. The solid lines
are calculated using equation \ref{eq:frequency} with the following
parameters: $M_{eff}=720$ emu.cm$^{-3}$, $\gamma=1.835\times10^{7}$
s$^{-1}$, $H_{u}=30$ Oe, $H_{b}=40$ Oe. Note that the $H_{me}$
is deduced from Figure \ref{Fig_dHres}.\textcolor{black}{{} The sketches
show how the magnetization direction rotates as function of the applied
voltage.}}
}

\textcolor{black}{\label{Fig_Intensity_variation} }
\end{figure}

\textcolor{black}{For this purpose, the pumping field was applied
parallel to $x$ direction ($\vec{h}_{rf}\parallel\vec{x}$ ), i.
e. perpendicular to the initial easy axis (see Figure \ref{Fig_Sketch_Actuator_Zero_Voltage_FMR}).
In order to ensure an initial uniform magnetization along the $y$
axis (initial easy axis), a small bias field of 40 Oe was applied.
In this condition and in absence of applied voltage, $\vec{M}$ is
supposed to be uniform and aligned along $y$ axis. Then, the voltage
applied to the actuator, in steps of 10 V, leads to the develop}ment
of an indirect magnetoelectric filed parallel to the $x$ axis, which
tends to align $\vec{M}$ along this axis. The normalized component
of the magnetization along the indirect magnetoelectric field $m_{x}(H_{me})$
is represented in Figure \ref{Fig_Intensity_variation} using the
following formula deduced from the model presented by Counil \textit{et
al.} \cite{Counil_JAP_2004}: 
\begin{equation}
m_{x}(H_{me})=\frac{\cos\left(\arcsin(\sqrt{A_{N}(V)}\right)}{\cos\left(\arcsin(\sqrt{A_{N}(150V)}\right)}\label{eq:amplitude}
\end{equation}

where $A_{N}(150V)$ and $A_{N}(V)$ are the normalized (by the 0V
amplitude spectrum) amplitude of the 150 V spectrum and of the applied
voltage spectra, respectively. The red solid line in Figure \ref{Fig_Intensity_variation}
corresponds to the theoretical value of the normalized $x$-component
of the magnetization as function of $V$. One can note that the rotation
is complete at around 40 V ($H_{me}=70$ Oe) where the magnetoelectric
field totally compensates the small bias field ($H_{b}=40$ Oe) and
the initial anisotropy field (30 Oe). The confrontation between the
model and the experimental data proves that a voltage induced magnetization
rotation has been performed. Another proof of the effective rotation
of the magnetization comes from the voltage dependence of the uniform
mode frequency as function of the applied voltage. The blue solid
line in Figure \ref{Fig_Intensity_variation} refers to the simulation
of this frequency being given the $H_{me}(V)$ dependence previously
determined (see Figure \ref{Fig_dHres}). This dependence highlights
the well-known softening of the frequency when the magnetization rotates
from a hard axis to an easy one \cite{Zighem_JPCM_2007}; the frequency
minimum directly gives the voltage at which the system is in-plane
isotropic (i. e. where $H_{me}$ compensates the initial anisotropy
field and the bias field). Both the experimental frequency and the
amplitude dependencies are well fitted by the solid lines and they
clearly show a 90 degree of the magnetization direction. The theoretical
values have been calculated by using a magnetic energy density of
the film where the indirect magnetoelectric field is modeled by a
voltage dependent uniaxial anisotropy field. In these conditions,
the following analytical expression of the frequency can be derived
from the Smit-Beljers equation \cite{Zighem_JAP_2014}:

\begin{flalign}
\frac{2\pi f}{\gamma} & =\sqrt{\left(H_{me}(V)-H_{u}\right)\cos2\varphi+H_{b}\sin\varphi}\times\nonumber \\
 & \sqrt{4\pi M_{s}+H_{b}\sin\varphi+H_{u}\sin^{2}\varphi+H_{me}(V)\cos^{2}\varphi}\label{eq:frequency}
\end{flalign}

where $\varphi$ is evaluated at each increment of the applied voltage.
Finally, it should be noted that the spectra profiles were not always
close to a Lorentzian one, especially at high applied voltages, which
make this modeling really robust \cite{Counil_JAP_2004,Counil_JAP_2005}. 

In summary, the indirect magnetoelectric coupling in artificial heterostructure
was used to rotate the magnetization direction. MS-FMR was used in
two different modes to: i) quantitatively measure the voltage-dependence
of the indirect magnetoelectric field (which is strain-induced by
the piezoelectric actuator) and to ii) observe the rotation of the
magnetization. The voltage dependence of the $H_{me}$ was found to
be almost linear (in the range 0V-200V) and was used during the frequency
and amplitude variations of the sweep-frequency spectra modeling.

\end{document}